\documentclass[aps,prl,twocolumn,showpacs,floatfix,superscriptaddress,amssymb,amsmath]{revtex4-1}
\usepackage{graphicx}
\usepackage{dcolumn}
\usepackage{bm}
\usepackage{color}
\newcommand{\be}{\begin{equation}}
\newcommand{\ee}{\end{equation}}
\newcommand{\bea}{\begin{eqnarray}}
\newcommand{\eea}{\end{eqnarray}}
\def\nn{\nonumber\\}

\begin{document}

\title{From adiabatic to non-adiabatic pumping in graphene nanoribbons}

\author{Tejinder Kaur} 
\affiliation{Department of Physics and Astronomy, Nanoscale and Quantum Phenomena Institute, and Condensed Matter and Surface Science Program, Ohio University, Athens, OH 45701, USA} 

\author{Liliana Arrachea} 
\affiliation{Departamento de F$\acute{i}$sica, Facultad de Ciencias Exactas y Naturales, Universidad de Buenos Aires, Pabell$\acute{o}$n I, Ciudad Universitaria, 1428 Buenos Aires, Argentina}  

\author{Nancy Sandler}
\affiliation{Department of Physics and Astronomy, Nanoscale and Quantum Phenomena Institute, and Condensed Matter and Surface Science Program, Ohio University, Athens, OH 45701, USA} 

 \date{\today}

 \begin{abstract} 
Non-equilibrium two-parameter pumping transport through graphene ribbons, attached to reservoirs is described. A tight-binding model is solved using Keldysh formalism, and the crossover between adiabatic and non-adiabatic regimes is studied. Pumped dc currents through armchair ribbons show properties common in two-dimensional systems. The width-dependent dc current in zigzag ribbons, reveals that edge states akin to those in two-dimensional topological insulators, do not contribute to pumped transport in the adiabatic regime. The interplay between propagating and evanescent modes is discussed.

 \end{abstract}

\maketitle

The study of transport mechanisms in quantum devices via dc currents generated by ac fields has become an active research area in recent years. Since the theoretical proposal by Thouless \cite{Thouless} -later developed for closed and open systems in several works \cite{Niu,Buttiker,Brouwer,Altshuler,Floquet,Moskalets}-, the application of a periodic perturbation to pump dc charge or spin current was achieved in various experimental settings \cite{Switkes_expt,nanotube_expt}. Most of these works focused on the adiabatic regime (low driving), in configurations with two or more periodically changing parameters that yield dc currents proportional to the driving frequency. 

New insights into pumping have been arising in studies on graphene \cite{Novoselov}. A two-parameter pumping model in the adiabatic regime, based in the Dirac Hamiltonian, obtained an enhanced pumped current (as compared with semiconductor materials), which was attributed to the unusual persistence of evanescent modes in the presence of Dirac points \cite{Elsa1}. Green's function methods, analyzed the effects of resonant tunneling in a similar configuration, and found anomalous behavior in this regime \cite{Grichuk}. Other works have analyzed pumping in bilayer systems \cite{Foa1} and mono parametric pumping \cite{Gu}. 

Among the extraordinary properties of graphene, there are those arising from confinement effects. It is known that finite sample conductances are strongly dependent on the boundaries \cite{Abanin,Akhmerov}. Studies on isolated graphene ribbons revealed the existence of unique features in the band-structure due to different edge terminations \cite{Nakada, MahdiNJP}. The most salient aspect is the remarkable width-dependence of the energy spectrum with gapped, gapless and highly quasi-degenerate bands as the terminations are changed. 

On the other hand, in experimental setups, ribbons are connected to metallic leads, a fact with significant consequences in the context of dc transport \cite{Schomerus}. The combined effects of finite sizes and contact to leads, pose new challenges in our understanding of non-equilibrium transport. The aim of the present work is to analyze the properties of quantum pumping in graphene ribbons, induced by two local ac gate voltages operating with a phase-lag, in a setup that includes the effect of metallic contacts. The analysis presented below includes a description of the adiabatic regime and the crossover to the non-adiabatic regime for different edge terminations, in a wide range of driving frequencies, voltage amplitudes and doping regimes. We show that dc pumped currents in armchair nanoribbons (ANR) exhibit properties common to other two-dimensional systems. In remarkable contrast, zigzag terminated ribbons (ZNR) of appropriate width, show a vanishing  pumped current  in the adiabatic regime. This is a manifestation of the subtle width-dependent band-structure of these systems. These zigzag ribbons are akin to two-dimensional topological insulators, making our results relevant for those materials also.

\begin{figure}[!]
 \includegraphics[width=0.4\textwidth]{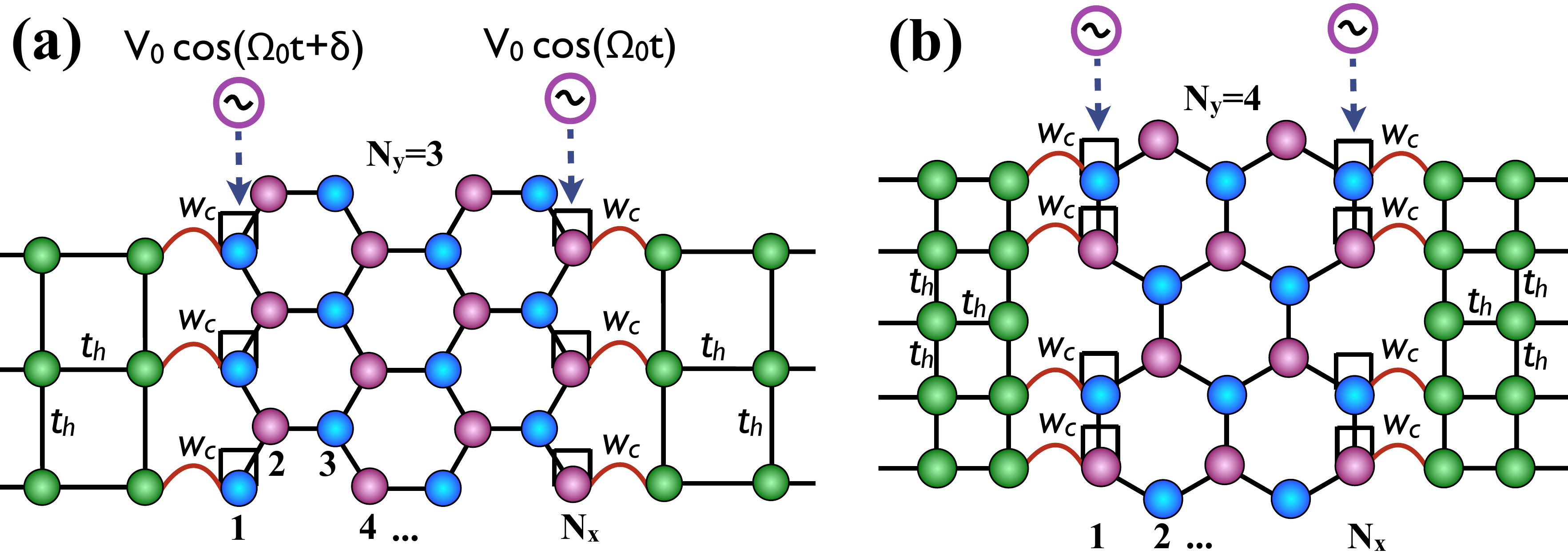}
 \caption{(a) Armchair ($N_{y}=3$; $N_{x}=8$) and (b) zigzag ($N_{y}=4; N_{x}=5$) graphene nanoribbons, attached to reservoirs and driven by time-dependent voltages with amplitude $V_0$, frequency $\Omega_0$ and phase difference $\delta$.}
 \label{arm_zig}
\end{figure}

{\it Theoretical treatment}-- We consider the setup of Figure ~\ref{arm_zig}, with a graphene ribbon along the $x-$direction attached to  metallic reservoirs. The ribbon's dimensions are given in terms of its width $W = N_y$ (number of rows) and  length $L = N_x$ (number of atoms) in multiples of the interatomic distance $a_{0} = 0.142 {\it nm}$. Two local ac dephased harmonic potentials with equal amplitudes drive the ribbon out of equilibrium. The system is described by the Hamiltonian: $H(t) = H_\text{rib}(t) + H_\text{res} + H_\text{tun}$, where:
\bea
H_\text{rib}(t)=- &t_{h}& \sum_{ <{\bf r},{\bf r}^{\prime}>}( c_{\bf r }^{\dagger} c_{\bf r^{\prime} } + h. c. ) +\;\;\;\;\;\;\;\;\;\;\;\;   \nn
\sum_{\alpha=L,R}\sum_{{\bf r}_{\alpha}} [E_{0}& +& V_0 \cos(\Omega_0 t + \delta_{\alpha})] c_{{\bf r}_{\alpha}}^{\dagger} c_{{\bf r}_{\alpha}}, 
\label{eq:ham1}
\eea
where $t_{h}$ is the hopping amplitude, $c_{{\bf r}}^\dagger (c_{\bf r})$ is the creation (destruction) operator in state ${\bf r} = 
(x, y)$, and the sum runs over nearest-neighbor sites only. Static barriers of height $E_0$ located at sites ${\bf r}_{L}$ and ${\bf r}_{R}$ separate the ribbon from the reservoirs. $V_{0}$, $\Omega_{0}$ and $\delta_{\alpha}$ correspond to the amplitude, frequency and phase lag between the two ac potentials applied on top of the barriers. For simplicity we take $\delta_{L}=0$ and $\delta_{R}=\delta$. A tight-binding Hamiltonian  for a square lattice with constant $a^{\prime}_0$ models two semi-infinite reservoirs with $N_{y}'$ chains and hopping $t_{r}$: $H_\text{res} = \sum_{\alpha} \sum_{{\bf k}_{\alpha}} \varepsilon_{{\bf k}_{\alpha} }c^{\dagger}_{{\bf k}_{\alpha}} c_{{\bf k}_{\alpha}}$. Here  ${\bf k}_{\alpha}= (k_{\alpha, x}, k_{\alpha, y})$, and $\varepsilon_{{\bf k}_{\alpha} } =- 2 t_r [\cos(k_{\alpha, x} a^{\prime}_0)+ \cos(k_{\alpha, y} a^{\prime}_0)]$. The connection between ribbon and reservoirs is given by $H_\text{tun} = \sum_{\alpha} \sum_{{\bf k}_{\alpha}, {\bf r}_{\alpha}} (w_{{\bf k}_{\alpha},{\bf r}_{\alpha} }c^{\dagger}_{{\bf k}_{\alpha}} c_{{\bf r}_{\alpha}}+ H. c. )$. In both reservoirs, $\alpha = R, L$, the chemical potential is the same, with $\mu = 0$ indicating the neutrality point. 

To obtain the dc current through the ribbon, we use the non-equilibrium Keldysh Green's function technique \cite{kel} as implemented in Refs.~\cite{Lili12, lilimos}. This formalism produces dc currents as functions of $\mu$, $V_{0}$ and $\Omega_{0}$, at and beyond adiabatic and/or perturbative regimes, including details of the band-structure of the ribbon and the contacts. The building block is the retarded Green' s function 
\bea
&& G^R_{{\bf r}, {\bf r^{\prime}}}(t,t^{\prime})  =  
- i \Theta(t-t^{\prime}) \langle \{c_{\bf r}(t), c^{\dagger}_{\bf r^{\prime}}(t^{\prime}) \}\rangle \nonumber \\
& &=  \sum_{n=-\infty}^{+\infty} e^{-i n\Omega_0 t}  \int \frac{d \omega}{2 \pi} e^{-i \omega (t- t^{\prime})} 
 {\cal G}_{{\bf r}, {\bf r^{\prime}}}(n ,\omega).
\eea
The dc current flowing from reservoir $\alpha$ through the ribbon is given by \cite{lilimos}:
\bea
&&J_{dc, \alpha}=  \frac{e}{h}\sum_{\beta=L,R} \sum_{n=-\infty}^{\infty} \int_{-\infty}^{\infty} d\omega
\mbox{Tr}\{\hat{\Gamma}_{\alpha} (\omega+n \Omega_0) \times \nn
&&\hat{\cal G}(n,\omega) \hat{\Gamma}_{\beta} (\omega)  \hat{\cal G}^{\dagger} (n,\omega)\}  \big[f_\beta(\hbar \omega)-f_\alpha(\hbar \omega+n\hbar \Omega_0)  \big],
\eea
 where $f_{\alpha}(\omega)$ is the Fermi distribution function of reservoir $\alpha$, $\hat{\Gamma}_{\alpha}(\omega)$ is the hybridization matrix (obtained after integration of the reservoir degrees of freedom) with elements  $\Gamma_{ {\bf r}_{\alpha},{\bf r^{\prime}}_{\alpha} }(\omega)= 2 \pi 
\sum_{ {\bf k}_{\alpha}} w_{{\bf r}_{\alpha},  {\bf k}_{\alpha}}  w_{{\bf k}_{\alpha},{\bf r^{\prime}}_{\alpha}} \delta(\omega-\varepsilon_{{\bf k}_{\alpha}})$. All calculations are carried out at zero temperature (finite temperature effects will be considered elsewhere) and hopping parameters are chosen such that $t_h=t_r$. Furthermore, we set $t_{h}$ as the unit of energy, choosing $E_0 =  t_{h}$ and explore the range $[0 \leq V_0 \leq E_{0}]$ for the pumping amplitude. We confirmed that maximum dc currents are obtained with $\delta= \pi/2$ \cite{Brouwer,Moskalets,Lili12} and this value for $\delta$ is fixed in the rest of the paper. 

\begin{figure}[!]
 \includegraphics[width=0.46\textwidth]{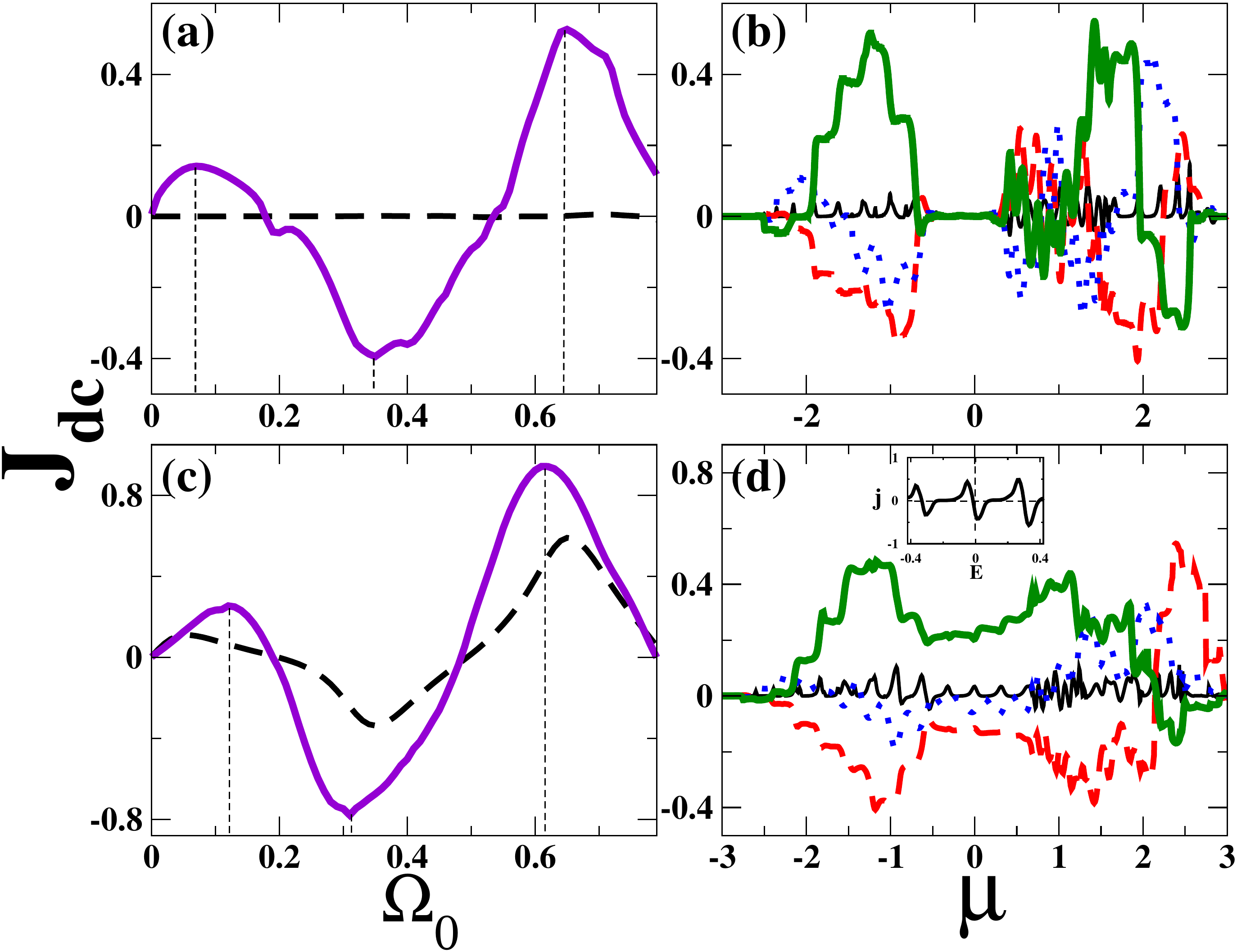}
 \caption{DC current as function of driving frequency $\Omega_0$ for ANR. (a) [(c)] Semi-conducting ($N_y=2$, $N_x=20$) [metallic ($N_y=4$, $N_x=20$)] ribbon. Solid (violet) and dashed (black) lines are for chemical potential $\mu=-1$ and $0$ respectively. Vertical dashed (black) lines indicate position of resonant frequencies. 
(b) [(d)] $J_{dc}$ as function of $\mu$ for semi-conducting [metallic] armchair. Thin solid (black), dashed (red), dotted (blue), and thick solid (green) lines correspond to $\Omega_{0} = 0.01, 0.3, 0.45$, and $0.6$ respectively. Inset shows current per unit energy vanishing at $\mu=0$. Other parameters are: barrier height $E_{0} = 1$, tunneling parameter $w_c = 1$, and pumping amplitude $V_0 = 0.2$. All energies are in units of $t_{h}$. The phase difference between potentials is set to $\delta= \pi /2$.}
\label{sqANR}
\end{figure}

{\it Armchair ribbon} - As shown in Figure~\ref{arm_zig}(a), these ribbons are connected to reservoirs through zigzag edges. Square and honeycomb lattices have $N_{y}' = N_{y}$ with equally spaced atoms. Panel $(a)$ [$(c)$] in Figure~\ref{sqANR} show results for the pumped current in semiconducting [metallic] ribbons, as a function of $\Omega_0$ for two representative values of $\mu$. Dashed lines are for $\mu = 0$, in the gap region for semiconducting ribbons [at the Dirac point in metallic ones].  As shown, the current vanishes for the frequencies explored in the semiconducting ribbon, remaining finite in the metallic ones.  For larger $\mu = -1.0 t_{h}$ (full lines), pronounced maxima and minima reveal the existence of discrete electronic levels in the structure. In metallic ribbons, (or semiconducting ones with chemical potential away from the gap), we identify the {\it adiabatic regime} by the linear increase of the current with $\Omega_{0}$ at low frequencies \cite{Brouwer, Moskalets}. For higher frequencies, the current displays maxima and minima as function of $\Omega_0$, indicating interference between different electronic levels \cite{Lili12}. Qualitatively similar results were obtained for various values of $\mu \neq 0$ (not shown).

Panels (b) and (d) show the  current versus $\mu$ for different values of $\Omega_{0}$. In the adiabatic regime  ($\hbar \Omega_{0} = 0.01 t_{h}$), the current vanishes for values of $\mu$ within the gap in semiconducting ribbons. For metallic ribbons distinct peaks appear, indicating the existence of  discrete energy levels that originate from the linear bands confined by the double barrier structure. Values for resonant frequencies could  be determined from this data by taking the difference between the positions of any two peaks. The peak at $\mu=0$ in metallic ribbons reflects the current transported by all modes up to $\mu=0$. The inset in panel (d) shows that the current per unit energy vanishes at the Dirac point because of the Klein paradox \cite{Schomerus,Elsa1}.  At larger and smaller values of $\mu$, charge transport occurs through many levels with various interference effects taking place. Notice that the contact with metallic leads plus static barriers produce a particle-hole asymmetric current. It is remarkable however, that for a given resonant frequency, the sign of the current in metallic ribbons is unchanged for a wide range of fillings.
 
{\it Zigzag ribbon} - The connection between ZNR and reservoirs requires to match lattices with different atomic spacings. We follow the approach used in Ref.~\cite{Schomerus}, as shown in Figure \ref{arm_zig}(b). The band-structure of zigzag ribbons exhibits a characteristic zero-energy band that is usually associated with the presence of edge states \cite{Nakada, MahdiNJP}. Several works have revealed width-dependent features in the band-structure with a dramatic effect on the conductance \cite{Akhmerov, DFTpaper}. For long ribbons it has been shown \cite{MahdiNJP} that the zigzag/anti-zigzag (or even/odd) effect is a manifestation of the different nature of degenerate states at the band center. For even values of $N_{y}$  the degeneracy is accidental while for odd values, it is protected by the underlying sub-lattice symmetry, being quite robust. 

\begin{figure}[!]
 \includegraphics[width=0.47\textwidth]{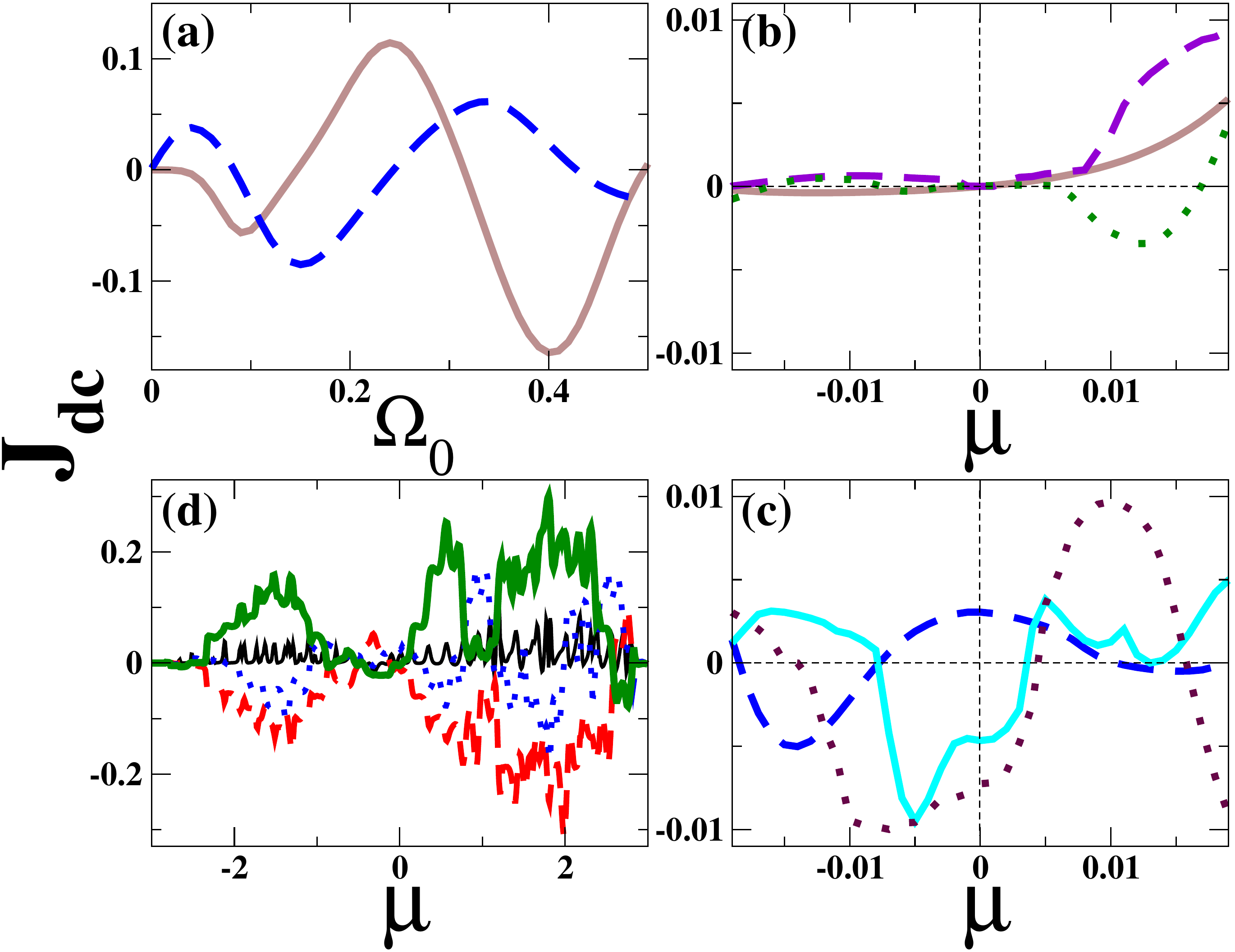}
 \caption{DC current as  a function of pumping frequency $\Omega_0$ at $\mu=0$ for ZNR. (a) Solid (brown): odd ($N_y=3$) chains; dashed (blue): even chains ($N_y=4$). (b) $J_{dc}$ as a function of $\mu$ in the adiabatic regime ($\hbar \Omega_0=0.01t_h$) for odd chains. Solid (brown), dashed (violet), and dotted (green) lines are for $N_y = 3, 5,$ and 7 respectively. (c) Same as (b) for even chains. Dashed (blue), solid (cyan) and dotted (maroon) lines are for $N_y=4, 6,$ and $8$ respectively. (d) $J_{dc}$ as a function of $\mu$ for even chains ($N_y=4$). Thin (black) solid, dashed (red), dotted (blue), and thick solid (green) lines for $\Omega_0 = 0.01, 0.2, 0.3,$ and $0.4$ respectively.
Other parameters are same as in Figure \ref{sqANR}.}
 \label{sqZNR}
\end{figure}

To gain insight into the nature of these peculiar states, we show in  Figure \ref{sqZNR}(a) results for the dependence of $J_{dc}$ on the driving frequency $\Omega_{0}$ for odd- (solid line) and even-chain (dashed line) ZNR at $\mu =  0$. Similar to metallic ANR, the current exhibits alternating sign changes and indications of resonant frequencies away from the adiabatic regime. Moreover, even-chain ribbons display the usual linear dependence characteristic of  small $\Omega_{0}$. This is consistent with the accidental nature of the degeneracy of the states at the center of the band, that is removed by the confining potential. As a consequence, these ribbons exhibit similar behavior to armchair ones. In contrast, odd- ZNRs exhibit no current {\it at all} at small frequencies, consistent with reported results in the resonant tunneling regime \cite{Grichuk}. This behavior is  more clearly exposed in the dependence of $J_{dc}$ with $\mu$ as shown in panels $(b)$ and $(c)$  for ribbons with different widths and fixed aspect ratios. Edge states (that persists even in the presence of metallic leads)  in odd-chain ribbons {\it do not contribute} to current transport in the adiabatic regime. 

An analysis of the current as a function of $\mu$ for different values of $\Omega_{0}$ is shown in panel $(d)$, in the case of an even-chain ribbon. Solid lines (adiabatic regime) describe the characteristic peak structure due to confinement effects. For resonant and non-resonant frequencies, the current changes sign more often as compared to metallic armchair ribbons, indicating the presence of quasi-degenerate states. 

In closing, the expected quadratic dependence of the dc current on the pumping amplitude for low amplitudes \cite{Lili12,pumping1dto2d}, was confirmed for all ANRs and ZNRs studied.

\begin{figure}[!]
 \includegraphics[width=0.47\textwidth]{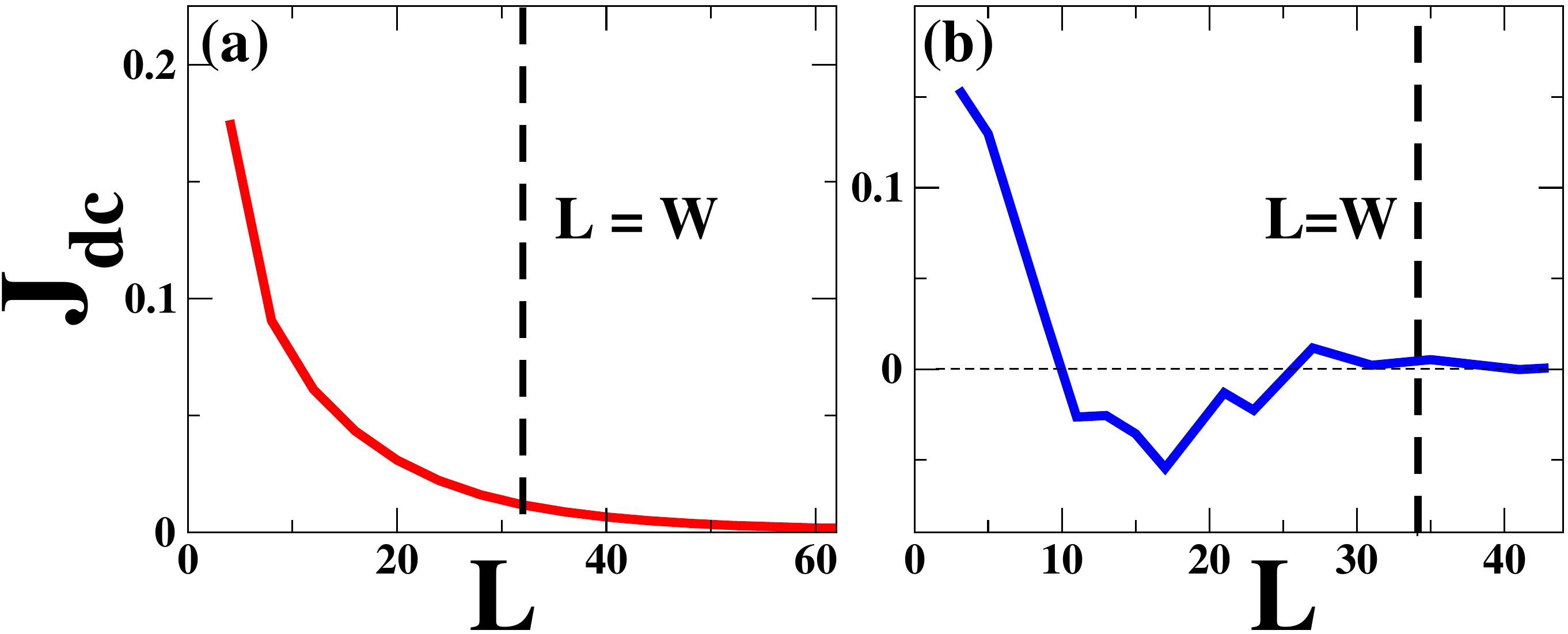}
 \caption{DC current as a function of length L (=$N_x$) at $\mu=0$ and $\Omega_0=0.01$ for (a) semi-conducting ANR ($N_y=14$) and (b) odd chain ZNR ($N_y=19$). Other parameters are the same as those in Figure \ref{sqANR}.}
\label{ev_modes}
\end{figure}

{\it Role of evanescent modes} - Up to here, our results characterize dc currents due to pumping through propagating modes. These are the dominant modes in transport when confined systems are long enough. Although confined graphene shows properties that strongly depend on boundary conditions, it is also possible to obtain a {\it universal} behavior in ribbons under the right conditions. The universality is evident when $ W \gg L$ and it explains the unique values of conductance and enhanced pumped currents in the adiabatic limit \cite{Elsa1,Schomerus,Tworzydlo}. For shorter and wider ribbons, transport is possible because of modes from the reservoirs, leaking through the barriers, to the system (evanescent). The crossover between these two regimes is described in Figure \ref{ev_modes}. Panel $(a)$ shows the dc current as a function of length for semi-conducting ANR (gap $\propto 1/W$) with the exponential decrease characteristic of evanescent modes.
Panel $(b)$ shows similar results for odd-chain ZNR with null current in the adiabatic regime. Thus  transport proceeds through evanescent  modes only with an exponentially small contribution at long lengths. While the qualitative behavior for $L \ll W$ and $L \gg W$ is similar to ANR, the
the current changes sign for intermediate lengths. This
 crossover provides further evidence of the existence of peculiar states in these systems. The  origin for the sign change  in the current can be understood as a consequence of the increasing number of quasi-degenerate edge states as $L$ increases, with the concomitant opening of new resonant frequencies. This means that, while $\Omega_{0}$ is not a resonant  for  a given
$L$, it can become resonant at a larger one. The edge modes that originally did not  transport current, become active in the pumping regime and compete with the evanescent modes already present. Although the picture presented here captures qualitatively the behavior of the current in this intermediate regime, a quantitative analysis
of the values of the parameters that determine the sign of the resulting current, implies further investigations of the detailed evolution of the spectrum with $L$.

{\it Conclusions}-- We provided a full description of quantum pumping in graphene nanoribbons with armchair and zigzag terminations connected to metallic leads, relevant for experimental settings. The phenomena observed in ANRs is characteristic of non-equilibrium transport in 2d systems with similar band-structure properties (i.e., metallic or semiconducting) \cite{pumping1dto2d}. In contrast, ZNRs exhibit a novel width-dependent pumped current that reflects the peculiar nature of edge states.  For appropriate ribbon widths, edge states akin to those existing in topological insulator materials have vanishing pumped dc current in the adiabatic regime. Furthermore, these modes, that persist in the presence of metallic contacts, compete with existent evanescent modes originating in the contacts. 
Although the models considered are within computationally tractable sizes (few nanometers), they correspond  to systems available in current experimental setups \cite{graphenegrowth}.  A simple extrapolation to realistic ribbon sizes, predicts dc pumped currents in the range $\mu A$ to $m A$ at $10-100 GHz$ resonant frequencies. Phenomena associated with the neutrality point should persist in a range of dopings ($\Delta \mu \ll t_h$) as well as at realistic values of static barriers and driving voltage amplitudes \cite{Switkes_expt}. In general, pumping involves heat dissipation. Within the weak driving regime, energy is dissipated at a rate $\propto \hbar \Omega_0^2$, while the pumped heat rate is $\propto k_B T \Omega_0$, being $T$ the temperature of the reservoirs \cite{liliheat}. An efficient operation of the pumping mechanism with low heating is, thus, expected, in the regime satisfying $\hbar \Omega_0 < k_B T$. 

{\it Acknowledgements}-- This work was supported partially by NSF SPIRE and MWN/CIAM grants. LA is supported by CONICET, MINCYT and Guggenheim Foundation.
The authors thank E. Prada, P. San-Jose, J. Schelter, and B. Trauzettel for discussions.

\end{document}